# Quantum Supremacy in Tomographic Imaging: Advances in Quantum Tomography Algorithms

Hyunju Lee[1] and Kyungtaek Jun[1,2,§]

[1] Quantum Research Center, QTomo, Chungcheongbuk-do, 28535, South Korea
[2] Chungbuk Quantum Research Center, Chungbuk National University, Chungcheongbuk-do, 28644, South Korea
[§] Corresponding author email: ktfriends@gmail.com

*Abstract*— **Quantum computing has emerged as a transformative paradigm, capable of tackling complex computational problems that are infeasible for classical methods within a practical timeframe. At the core of this advancement lies the concept of quantum supremacy, which signifies the ability of quantum processors to surpass classical systems in specific tasks. In the context of tomographic image reconstruction, quantum optimization algorithms enable faster processing and clearer imaging than conventional methods. This study further substantiates quantum supremacy by reducing the required projection angles for tomographic reconstruction while enhancing robustness against image artifacts. Notably, our experiments demonstrated that the proposed algorithm accurately reconstructed tomographic images without artifacts, even when up to 50% error was introduced into the sinogram to induce ring artifacts. Furthermore, it achieved precise reconstructions using only 50% of the projection angles from the original sinogram spanning 0° to 180°. These findings highlight the potential of quantum algorithms to revolutionize tomographic imaging by enabling efficient and accurate reconstructions under challenging conditions, paving the way for broader applications in medical imaging, material science, and advanced tomography systems as quantum computing technologies continue to advance.**

*Index Terms*— **Quantum tomography, quantum computing, quantum optimization tomographic reconstruction, quantum supremacy, artifact removal.**

## I. INTRODUCTION

QUANTUM computing has emerged as a revolutionary paradigm, surpassing classical methods in solving complex computational challenges. At the heart of this field lies the concept of quantum supremacy, which denotes the milestone where a quantum computer executes a calculation that classical systems cannot feasibly complete within a reasonable timeframe. Introduced by John Preskill in 2012, quantum supremacy marks a pivotal breakthrough, showcasing the ability of quantum processors to solve problems significantly faster than their classical counterparts [1]. Demonstrations of quantum supremacy, including integer factorization [2], boson sampling [3], quantum random circuit sampling [4], [5], and the Quantum Approximate Optimization Algorithm (QAOA) [6], not only validate quantum advantage but also propel advancements in quantum optimization techniques as quantum hardware continues to evolve.

Beyond theoretical milestones, practical applications underscore quantum computing's transformative potential across diverse domains. For instance, quantum computers excel in simulating the electronic structures of molecules. In 2019, Google's Sycamore quantum processor computed the energy of quantum systems [7], a significant breakthrough in quantum chemistry with profound implications for drug discovery and materials science [8]. Similarly, quantum computing has demonstrated promise in addressing complex combinatorial optimization problems, which are critical in industries such as logistics, finance, and manufacturing. Volkswagen's implementation of quantum computing to optimize traffic flow exemplifies its capability to handle large-scale optimization tasks efficiently, leading to cost reduction and operational improvements [9].

Quantum algorithms are also poised to revolutionize data processing and analysis. Quantum machine learning approaches, such as the quantum k-nearest neighbors (QkNN) algorithm [10], exhibit superior speed and accuracy in data classification compared to classical methods, enhancing efficiency in applications like financial modeling [11] and image recognition [12]. In cryptography, Shor's algorithm underscores the immense computational power of quantum systems by posing a direct threat to widely used encryption protocols such as RSA encryption [13]. This development has driven extensive research into post-quantum cryptography to safeguard digital security against quantum threats. The financial sector has also begun leveraging quantum computing for tasks such as portfolio optimization, risk analysis, and accelerated Monte Carlo simulations. Companies like IBM and JP Morgan have explored quantum computing for option pricing, unlocking new possibilities for precise financial modeling and enhanced investment strategies [14].

These advancements extend beyond demonstrating quantum supremacy, illustrating the broader potential of quantum computing to redefine problem-solving paradigms across science, technology, and industry. As quantum hardware matures, its profound impact on computation is poised to establish it as a cornerstone of future innovation.

The concept of quantum supremacy is increasingly being explored across diverse fields, including tomographic reconstruction, offering new possibilities for faster and more accurate image processing. Motivated by the need for high-speed computations and enhanced reconstruction accuracy, we investigate the potential of quantum computing in this domain. The Quadratic Unconstrained Binary Optimization (QUBO) formulation serves as the foundation for solving linear equations [15] and eigenvalue problems [16], demonstrating remarkable computational efficiency, particularly when implemented on quantum annealers optimized for such tasks. Building on this framework, we develop a quantum algorithm for tomographic image reconstruction by formulating the QUBO problem based on sinogram pixel values. Experimental results validate the efficacy of this approach, highlighting the applicability of quantum algorithms in advancing tomographic image reconstruction.

Jun initially proposed a quantum optimization algorithm for tomographic image reconstruction [17]. Expanding on this work, Jun and Lee addressed the challenge of limited qubits in logical quantum systems by incorporating a superposition step using mass attenuation coefficients (MAC). This enhancement facilitated simultaneous tomographic image reconstruction and segmentation [18], marking a significant advancement in quantum imaging techniques.

Quantum supremacy has already demonstrated transformative potential across various disciplines, and tomographic imaging stands to benefit significantly from these developments. As a crucial tool in medical diagnostics, tomographic imaging requires precise and rapid reconstruction to ensure accurate clinical assessments. However, multiple factors, including motion artifacts, can degrade image quality and complicate interpretation. Previous research, including contributions by Jun, has sought to mitigate such artifacts by developing alignment methods that satisfy the Helgason-Ludwig consistency condition, enabling high-quality imaging even in cases of rotational misalignment [19]–[22]. Despite these advancements, challenges such as ring artifacts and constraints in projection angles remain unresolved. Hence, this study introduces a quantum optimization algorithm specifically designed to address these limitations, demonstrating its potential to enhance tomographic image reconstruction.

Building on this foundation, this paper presents one of the first efforts to integrate quantum supremacy into tomographic reconstruction. The proposed quantum algorithm effectively tackles critical challenges in reconstructing tomographic images from raw data, particularly in scenarios with limited projection angles and high noise levels. By leveraging quantum optimization, our approach offers a groundbreaking method for achieving faster and more accurate tomographic reconstructions, ultimately improving diagnostic quality. To assess the efficacy of the algorithm, we conducted extensive experiments comparing classical and quantum reconstruction techniques. These experiments encompassed cases with noisy projection data, induced ring artifacts, and scenarios simulating the missing wedge problem in electron tomography by restricting projection angles [23]. The results highlight the broad applicability of our algorithm, demonstrating its relevance not only in medical imaging but also in various tomography applications.

The proposed quantum tomographic algorithm is adaptable to all types of beam sources, leveraging the QUBO model to represent detector pixel values and the internal structure of the sample under analysis. For rigorous validation, this study employs a parallel-beam source, ensuring precise experimental verification and clarity in demonstrating the effectiveness of the algorithm.

The key contributions of this study are as follows:
1. **Development of a Quantum Tomographic Image Reconstruction Algorithm:** We propose a novel quantum algorithm capable of accurately reconstructing tomographic images, even under challenging conditions such as incomplete sinogram data and significant error injection. This approach demonstrates quantum supremacy by outperforming classical reconstruction methods in both speed and accuracy.
2. **Enhanced Robustness Against Artifacts:** Utilizing quantum optimization techniques, our algorithm effectively mitigates ring artifacts and reconstructs high-fidelity images despite sinograms containing up to 50% error. This performance significantly surpasses that of simulated annealing (SA) and classical methods such as Fast Fourier Transform (FFT).
3. **Reduction in Required Projection Angles:** Our method achieves high-quality tomographic image reconstruction using only 50% of the projection angles (from 0° to 180°), addressing critical challenges related to limited projection data and missing wedges in tomography systems.

The remainder of this paper is structured as follows: Section 2 provides the foundational background, including an overview of the quantum tomographic image reconstruction algorithm (Section 2.1) and the quantum tomographic image segmentation algorithm (Section 2.2). Section 3 details the methodology, covering the experimental data (Section 3.1), experimental settings (Section 3.2), and the tomographic image reconstruction approach using error-excluded sinograms (Section 3.3). Section 4 presents the results, comparing our method to classical reconstruction techniques (Section 4.1) and other classical optimization methods (Section 4.2), with a focus on ring artifact removal (Sections 4.1.1 and 4.2.1) and reconstruction with limited projection angles (Sections 4.1.2 and 4.2.2). Finally, Section 5 provides a comprehensive discussion and conclusion.

## II. BACKGROUND

Tomographic imaging involves rotating a sample and capturing projection images using a parallel-beam source at multiple angles. These projections are mathematically processed through the Radon transform, which computes integral values along specific lines within the object. The resulting data is sequentially arranged according to the projection angles, forming a sinogram—a crucial intermediate representation used to reconstruct the original tomographic image.

To achieve tomographic image reconstruction, the pixel values of the target image are first encoded as qubits. The reconstruction process then minimizes the discrepancy between the sinogram obtained from the actual object and the sinogram computed from the superposed tomographic image. This discrepancy serves as a cost function, formulated within the

QUBO model, as expressed in (1).

$$f(\vec{q}) = \sum_{i=1}^{n} Q_{i,i} q_i + \sum_{i<j}^{n} Q_{i,j} q_i q_j \quad (1)$$

where $\vec{q}$ denotes an element of $\mathbb{B}^n$. By solving for the minimum of the QUBO formulation, the tomographic image is reconstructed efficiently and accurately. The QUBO formulation utilizes binary variables (0 and 1), whereas the Ising model employs spin variables with values of −1 and 1. These two formulations are linearly transformable, enabling seamless interchange depending on the computational framework.

$$\sigma \to 2q - 1 \text{ or } q \to \frac{1}{2}(\sigma + 1) \quad (2)$$

where $\sigma_i \in \{-1, 1\}$ denotes a variable of the Ising model.

Quantum annealers and gate-based quantum computers utilize QUBO and Ising models as fundamental frameworks for quantum optimization. In quantum annealing, the objective is to find the minimum value of the model, whereas gate-based quantum computers, such as those implementing the QAOA, aim to maximize the value of the model. This study focuses on applying QUBO modeling to solve minimization problems, building upon our previous works [17], [18], which established the foundation for the proposed method. The key details from these studies are summarized below.

*A. Quantum Tomographic Image Reconstruction Algorithm*

In the initial study, Jun formulated the QUBO model for reconstructing tomographic images [17]. The pixel value of a projection image obtained at a given projection angle $\theta$ represents the intensity of X-rays transmitted through the thickness of the sample. In regions where the Beer-Lambert law applies, the pixel value at the $s$-th position in the projection image, denoted as $P(\theta, s)$ for a specific axial level and projection angle $\theta$, can be mathematically expressed as shown in (3):

$$P(\theta, s) = \int_{s}^{s+1} M(x, y) dl \quad (3)$$

Here, $M(x, y)$ represents the X-ray MAC at the $(x, y)$-coordinate of the sample, while the $l$-axis is aligned perpendicular to the X-ray direction. Now, consider a reconstructed image $I$. If $I$ consists of integer pixel values and the maximum pixel value is less than $2^{m+1}$, then each pixel in the reconstructed image can be represented as a combination of qubit variables:

$$I_{ij} \approx \sum_{k=0}^{m} 2^k q_k^{ij} \quad (4)$$

Here, $q_k^{ij}$ is a binary variable (0 or 1), and $I_{ij}$ can take any integer value from 0 to $2^{m+1} - 1$.

To process X-ray projection data obtained using a parallel-beam setup, the Radon transform is applied to the tomographic image $I$, generating a superposed sinogram $IP$. For a given projection angle $\theta$, the $s$-th position of $IP$ is computed as:

$$IP(\theta, s) = \sum_{i,j} c_{ij} I'_{ij} \quad (5)$$

where $I'_{ij}$ represents the pixel affecting $IP(\theta, s)$, and $c_{ij}$ denotes the overlapping area during projection.

The QUBO model for each pixel in the sinogram can be then computed as follows:

$$\{(IP - P)(\theta, s)\}^2 = \left\{ \sum_{i,j} c_{ij} I'_{ij} - P(\theta, s) \right\}^2 \quad (6)$$

$$= \left( \sum_{i,j} c I'_{ij} \right)^2 - 2P(\theta, s) \left( \sum_{i,j} c_{ij} I'_{ij} \right) + \{P(\theta, s)\}^2 \quad (7)$$

$$= \left( \sum_{i,j} c_{ij} \sum_{k=0}^{m} 2^k q_k^{ij} \right)^2 - 2P(\theta, s) \left( \sum_{i,j} c_{ij} \sum_{k=0}^{m} 2^k q_k^{ij} \right) + \{P(\theta, s)\}^2 \quad (8)$$

$$= \sum_{i,j} \sum_{k=0}^{m} 2^{2k} c_{ij}^2 q_k^{ij} + \sum_{i,j} \sum_{0 \le k < k' \le m} 2^{k+k'+1} c_{ij}^2 q_k^{ij} q_{k'}^{ij} + \sum_{\substack{i,i',j,j' \\ i \ne i' \text{ or } j \ne j'}} \sum_{0 \le k < k' \le m} 2^{k+k'+1} c_{ij} c_{i'j'} q_k^{ij} q_{k'}^{i'j'} - 2P(\theta, s) \left( \sum_{k} a_k \sum_{k=0}^{m} 2^k q_k^{ij} \right) + \{P(\theta, s)\}^2 \quad (9)$$

By using $\left(q_k^{ij}\right)^2 = q_k^{ij}$, we can obtain the first term of (8) as the sum of linear and quadratic terms. In (9), the first and fourth terms represent the linear terms, while the second and third terms correspond to the quadratic terms. Excluding the constant term, the remaining terms form the QUBO formulation for a single pixel in the sinogram. By summing these formulations across all pixels in the sinogram, the final QUBO model is derived. Therefore, the overall QUBO model for tomographic image reconstruction is calculated as follows:

$$\sum_{\theta=0}^{180-d\theta} \sum_{s=1}^{n} \{(IP - P)(\theta, s)\}^2 - \sum_{\theta=0}^{180-d\theta} \sum_{s=1}^{n} \{P(\theta, s)\}^2 \quad (10)$$

where $\theta$ denotes the projection angle, $s$ denotes the position of the sensor, and $d\theta$ denotes the amount of change in the projection angle. Here, the first term always takes a non-negative value, which implies that the minimum value of the

QUBO model is: $-\sum_{\theta=0}^{180-d\theta}\sum_{s=1}^{n}\{P(\theta,s)\}^2$. Furthermore, it can be observed that if a solution satisfies $IP = P$, the value of the QUBO model reaches this minimum.

### B. Quantum tomographic Image Segmentation Algorithm

Building on the initial formulation, Jun and Lee proposed an enhanced model to address the limitation of insufficient qubit availability by introducing a representation of tomographic image pixel values based on the MAC [18]. This approach also enabled the segmentation of tomographic images through a QUBO formulation, providing a comprehensive framework for improved image reconstruction and analysis. The following overview is adapted from their work.

Assume the sample exists in 3-dimensional space. Let $\alpha = \mu/\rho$ represent the X-ray MAC [24], and let $m$ denote the number of possible X-ray MAC values for each grid space. Each pixel $I_{ij}$ in the reconstructed 2D image can then be expressed as a combination of qubits and binary values:

$$I_{ij} = \sum_{k=1}^{m} \alpha_k q_k^{ij} \tag{11}$$

where $q_k^{ij}$ denotes a binary variable (0 or 1), and $I_{ij}$ takes values from the set $\{0, \alpha_1, \alpha_2, \cdots, \alpha_m\}$. In 3D space, the pixel position is represented as $(i, j, z)$, where $z$ denotes the axial level. Subsequently, by substituting (11) into (5), $IP(\theta, s)$ is expressed as a combination of MAC and qubits.

The QUBO model is then derived by applying the least squares method to minimize the difference between $P(\theta, s)$ and $IP(\theta, s)$.

$$\left(IP(\theta,s) - P(\theta,s)\right)^2 = \left(\sum_{i,j} c_{ij} I'_{ij} - P(\theta,s)\right)^2 \tag{12}$$

$$= \left(\sum_{i,j} c_{ij} \sum_{k=1}^{m} \alpha_k q_k^{ij} - P(\theta,s)\right)^2 \tag{13}$$

$$= \left(\sum_{i,j} c_{ij} \sum_{k=1}^{m} \alpha_k q_k^{ij}\right)^2 - 2P(\theta,s) \sum_{i,j} c_{ij} \sum_{k=1}^{m} \alpha_k q_k^{ij} + \{P(\theta,s)\}^2 \tag{14}$$

$$= \sum_{i,j} \sum_{k=1}^{m} c_{ij}^2 \alpha_k^2 q_k^{ij}$$
$$+ \sum_{i,j} \sum_{1 \le k < k' \le m} 2 c_{ij}^2 \alpha_k \alpha_{k'} q_k^{ij} q_{k'}^{ij}$$
$$+ \sum_{\substack{i,i',j,j' \\ i \ne i' \, or \, j \ne j'}} \sum_{1 \le k, k' \le m} 2 c_{ij} c_{i'j'} \alpha_k \alpha_{k'} q_k^{ij} q_{k'}^{i'j'} \tag{15}$$
$$- 2P(\theta,s) \sum_{i,j} c_{ij} \sum_{k=1}^{m} \alpha_k q_k^{ij} + \{P(\theta,s)\}^2$$

In this case, the final QUBO model is constructed by summing the above QUBO formulations across all pixels in the sinogram.

However, to achieve more stable results, an alternative approach can be employed to represent the pixel values of the tomographic image. To constrain the range of $I_{ij}$ to values between 1 and $\alpha_m$, the following representation can be used:

$$J_{ij} = \alpha_1 + \sum_{k=2}^{m} (\alpha_k - \alpha_{k-1}) q_k^{ij} \tag{16}$$

The QUBO model for $J_{ij}$ is derived by replacing $\alpha_2$ to $\alpha_m$ in the QUBO model of $I_{ij}$ with $(\alpha_2 - \alpha_1)$ to $(\alpha_m - \alpha_{m-1})$. By modifying the method of combining the reconstructed image's pixel values with qubits, this approach enhances the solution-finding performance on a quantum computer.

## III. METHOD

This study builds upon the methods introduced in our previous works [17], [18], which proposed effective frameworks for X-ray tomographic image reconstruction and segmentation using the X-ray MAC and QUBO formulations. In this work, we introduce restricted domain ranges to eliminate erroneous data, thereby enhancing the accuracy and reliability of the reconstruction process. The following sections outline the experimental setup and methodology used to validate this refined approach.

### A. Experimental Data

In the quantum algorithm for tomographic image reconstruction, experiments were conducted using various datasets to demonstrate quantum supremacy. However, if the pixel values of tomographic images span too wide a range or involve an excessive number of possibilities, representing these values with qubits would require an impractically large number of qubits. Given the limited number of qubits available on current quantum computers, we reduced the size of the image data and adjusted the range and number of pixel values for the experimental setup. A total of four datasets were used in the experiments, and their details are as follows:

*1) Shepp-Logan phantom image*

This dataset, shown in Fig. 1a, was generated using the 'shepp_logan_phantom' function from the scikit-image library, resized to a 50 × 50 resolution, and normalized so that all pixel values were set to 1 for experimental purposes.

*2) Artificial molar*

An X-ray image of a virtual molar tooth was captured using the Beamline 6C BioMedical Imaging facility at the Pohang Accelerator Laboratory, with an original resolution of 2560 × 2160 pixels. Detailed preprocessing steps and data acquisition methods can be found in [18]. Two experimental datasets were created for analysis: one resized to 100 × 100 pixels at axial level 27 and another resized to 50 × 50 pixels at axial level 15. Projection images were processed into sinograms, and pixel values were adjusted so that the tooth region was set to 1 and the background to 0, ensuring clear segmentation for the experiments.

*3) Body tomographic image from Kaggle*

A tomographic image from the SIIM Medical Images dataset [25] was used in this experiment for analysis. The original tomographic image had a size of 512 × 512 pixels, with pixel values ranging from 0 to 1918. To prepare the image for analysis, it was resized to 50 × 50 pixels, and each pixel was adjusted to have integer values ranging from 0 to 3. Non-relevant regions, not containing any objects, were removed before conducting the experiment.

*4) Head tomographic image from Kaggle*

A tomographic brain scan image from the dataset 'Computed Tomography (CT) of the Brain' [26] was used in this experiment. The brain tomographic image had a size of 512 × 512 pixels, with pixel values ranging from 0 to 255. For the purpose of the experiment, the image was resized to 30 × 30 pixels, and each pixel was adjusted to have integer values ranging from 0 to 10.

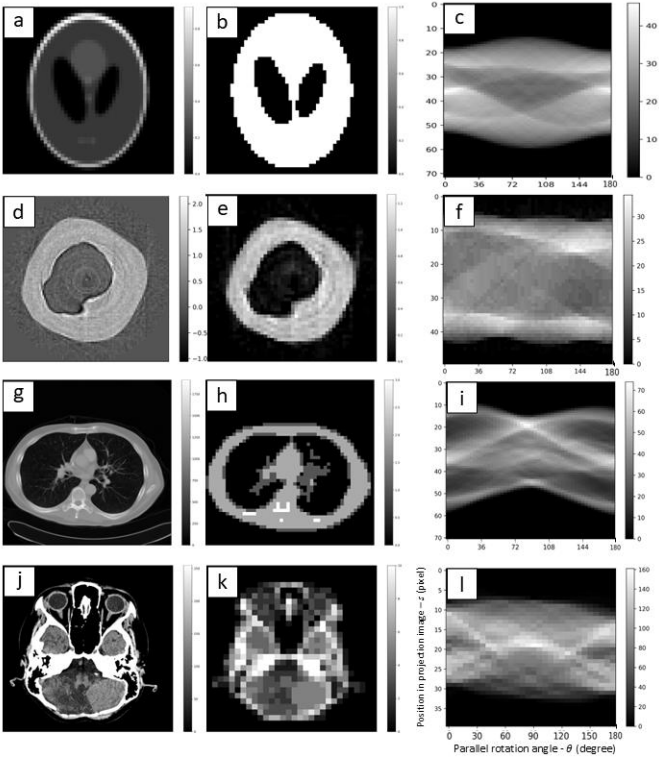

Fig. 1. Sinograms and corresponding tomographic images for the four datasets used in the experiments. Column 1: High-resolution tomographic images. Column 2: Low-resolution tomographic images derived from the first column. Column 3: The sinogram corresponding to the low-resolution tomographic images used in the experiment.

The sinogram preparation method used in our algorithm is consistent across all datasets except for the second one. In the first, third, and fourth rows of Fig. 1, the sinograms were generated from existing tomographic images. These images were resized and resampled to adjust the image size and pixel dimensions, followed by the application of the Radon transform. The transformation was performed using projection angles ranging from 0° to 180° (or 360°), with the number of projection angles set equal to the length of one side of the image.

In contrast, the second dataset was derived from X-ray data obtained by imaging an artificial tooth in 0.5° increments over a total of 360 projections. The X-ray data was then binned to the desired size, and a projection image from a predefined axial level was extracted to generate the sinogram. However, due to varying noise levels that may occur at different angles during the scanning process, the projection images may not connect smoothly across each angle. Therefore, preprocessing is required to mitigate the discontinuities in the sinogram and ensure smoother connections.

The first step of the preprocessing involves normalizing the pixel values of the sinogram based on the values from the first projection image. Since the X-ray mass attenuation coefficient ($\alpha$) of a single material is constant, we scale the entire sinogram to set $\alpha$ equal to 1, simplifying the QUBO formulation. The scaled sinogram is obtained by applying a classical reconstruction algorithm to the sinogram from the first step, reconstructing the image, and then multiplying the sinogram by the average pixel value of the material region. This average value is calculated by dividing the sum of the pixel values by the number of pixels in the region. In our experiments with actual data, we created and used the scaled sinogram following this two-step preprocessing process.

To conduct experiments on sinograms with errors, we introduced artificial noise into the prepared sinograms. The direct use of sinograms that inherently contain errors was not feasible due to the following limitation: the current constraints on the number of qubits available for quantum computing necessitate reducing the size of the sinogram. In real data, sinograms with errors typically exhibit defects within 10 lines of a large-sized image. However, during the resizing process, these localized errors blend with surrounding pixels, making them less distinguishable. Consequently, introducing artificial errors into the resized sinograms was necessary to ensure their suitability for experimental analysis. Hence, a single random position along the position axis of the sinogram was selected, and artificial errors were added by multiplying all the pixels at that position by a specific factor. This process was repeated at multiple random positions, with each position assigned a different factor, thereby introducing diverse error patterns into the sinograms. When these sinograms, containing artificial errors, were processed using FFT for tomographic image reconstruction, ring artifacts were observed.

Additionally, to simulate scenarios where data from specific projection angles are unavailable, we generated sinograms with incomplete projection data. While projection images are typically acquired over a range of 0°–180° (or 360°), we created sinograms using narrower angular ranges or by omitting projections at irregular intervals. These incomplete sinograms were also included in our experiments.

### B. Experimental settings

This study formulates the tomographic image reconstruction problem as a QUBO model and solves it using quantum sampling techniques. Specifically, we employed the quantum annealing system provided by D-Wave. Among the solvers available through D-Wave Leap, we utilized the hybrid binary quadratic model solver, `hybrid_binary_quadratic_model_version2p`. This solver can handle up to two million variables, making it well-suited for problems requiring a large number of qubits, such as ours.

For comparative analysis, we also solved the same QUBO model using classical computing methods: Gurobi (version 12.0.0), a state-of-the-art mathematical optimization solver, and SA, implemented using the D-Wave `SimulatedAnnealingSampler()` command. However, Gurobi requires an excessive amount of time to find a solution for problems with a large number of variables, as in our experiment. Therefore, we focused on the solution obtained from the presolve stage provided by Gurobi. This stage simplifies the optimization problem before the main solving process by reducing the problem size and complexity.

Moreover, we implemented the classic tomographic image reconstruction method—FFT—using the `iradon` function available in the scikit-image library (version 0.24.0). The FFT is a widely used algorithm for efficiently computing the inverse Radon transform, which reconstructs the tomographic image from projection data. Since the tomographic image reconstructed using FFT contains floating-point pixel values, we converted the pixel values to integers by rounding them to ensure a fair and intuitive comparison with other reconstruction methods that produce tomographic images with integer pixel values.

### C. Tomographic image reconstruction method using error-excluded sinograms

In previous studies, QUBO formulations were constructed for all pixels in the sinogram. In contrast, the method proposed in this work identifies regions in the sinogram affected by errors and excludes these regions from the entire domain. The QUBO formulation is then generated only for the unaffected regions, facilitating the reconstruction of the tomographic image.

Given a sinogram $P(\theta, s)$ defined over a domain $D$, the pixel value $I_{ij}$ is formulated as described in Equations (4), (11), and (16). If the error-affected region of the sinogram $P$ is denoted as $E$, the QUBO model for tomographic image reconstruction, as expressed in (17), is constructed by excluding $E$ from the domain $D$.

$$\sum_{(\theta,s)\in D\setminus E} \{(IP - P)(\theta, s)\}^2 - \sum_{(\theta,s)\in D\setminus E} \{P(\theta, s)\} \quad (17)$$

The resulting formulation focuses exclusively on the unaffected areas of the sinogram, ensuring a more robust and efficient reconstruction process.

The method for applying our algorithm to a sinogram with errors that cause ring artifacts is as follows. Errors that occur perpendicularly to the position axis of the sinogram, which can lead to ring artifacts, are detectable by examining variations in the density of the projection image. As illustrated in Fig. 2a, these errors can be identified by locating areas along the same position where pixel values across projection angles deviate significantly from surrounding values. Specifically, the detection process involves analyzing the density changes between adjacent pixels for each projection angle. By identifying positions where pixel values differ markedly due to errors, the affected regions can be pinpointed. This process is repeated across all projection angles, and the average is computed to determine rows with abrupt density changes, indicating the presence of errors.

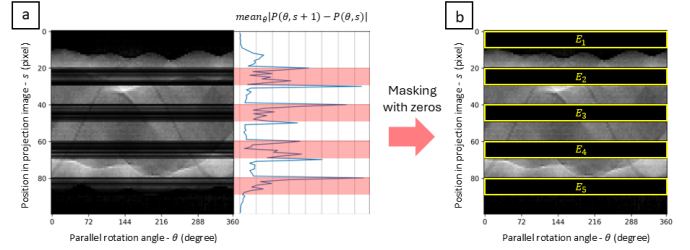

Fig. 2 Illustration of horizontal error regions in the sinogram that can induce ring artifacts in classical tomographic images, along with the process of identifying these error regions for algorithm application. (a) Sinogram with errors introduced at 10 positions, spaced 10 units apart along the position axis, and a graph depicting the mean absolute difference in density variation across angles. (b) Sinogram with error-containing regions masked to zero.

To apply this to our algorithm, the red regions identified in Fig. 2a were labeled as $E_1, E_2, E_3, E_4,$ and $E_5$. The error-affected region of the sinogram was then defined as $= E_1 \cup E_2 \cup E_3 \cup E_4 \cup E_5$. By excluding the error-affected region $E$ from the entire domain and generating the QUBO model, this approach is equivalent to using the QUBO model based on the sinogram where region $E$ is zero-masked, as shown in Fig. 2b. The Results section will present this case, along with various other experiments, to demonstrate the quantum supremacy achieved by our approach.

## IV. RESULTS

In this study, we compared the performance of the traditional FFT-based tomographic image reconstruction method with our proposed quantum algorithm to demonstrate quantum supremacy. Subsequently, we extended our approach to generate QUBO models for tomographic image reconstruction using additional datasets. By solving these QUBO models and comparing the results obtained through quantum computing with those from classical methods, we further validated the superiority of the quantum algorithm in tomographic image reconstruction.

### A. Comparison with the classical reconstruction method

#### 1) Reconstruction Results for Sinograms with Ring Artifact-Inducing Errors

We artificially introduced ring artifacts into our sample to simulate real-world data imperfections. The experiments were conducted under controlled conditions to accurately assess the impact of these artifacts on the algorithm's performance. Fig. 2a illustrates the sinogram used in the experiment to artificially generate ring artifacts. The images represent a $100 \times 100$ cross-section of a tooth sample at the $27^{th}$ axial level, with projection angles divided into 100 intervals from $0°$ to $360°$. To create the artificial artifacts, a specific random ratio was applied at every 10 positions along the sinogram, with a spacing of 10 rows. Figs. 3a and 3b show the tomographic images reconstructed from the sinograms in Figs. 2a and 2b, respectively, using FFT. To highlight the ring artifacts, the resulting image was constrained to pixel values between 0 and 1. Both Fig. 3a and Fig. 3b reveal challenges in clearly distinguishing the internal structures due to the presence of these artifacts.

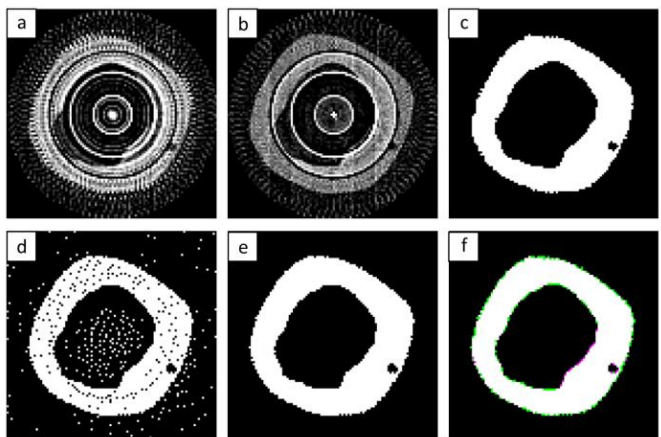

Fig. 3. Tomographic images reconstructed from the sinogram in Fig. 2 using our algorithm and FFT, for reconstructing a 100 × 100 cross-sectional image of the tooth sample. (a) Tomographic image reconstructed from the sinogram in Fig. 2a using FFT. (b) Tomographic image reconstructed from the sinogram in Fig. 2b using FFT. (c) Tomographic image obtained by applying FFT to the original sinogram before introducing artificial errors, followed by rounding using a threshold technique. (d) Tomographic image reconstructed from the sinogram in Fig. 2b using our algorithm. (e) Post-processed image from (b), with black and white dots removed. (f) Overlay of the FFT-based image in (c) and our post-processed result in (e).

As outlined in the previous section, our algorithm identified the error-affected regions in the sinogram as $E$ and generated the QUBO model within the domain excluding $E$. Solving this QUBO model using D-Wave's hybrid solver produced results similar to those shown in Fig. 3d, where the major structures are reconstructed with greater accuracy compared with the FFT-based reconstruction in Fig. 3b. However, several visible black and white dots persist. To generate a cleaner tomographic image, post-processing methods based on pixel connectivity were applied to fill black holes and remove white dots, resulting in the image shown in Fig. 3e. After post-processing, the final image presents a well-segmented cross-section of the tooth. Fig. 3f presents an overlay of the FFT result from the error-free original sinogram and the reconstructed image from our quantum algorithm, where 50% of the original sinogram data was removed. The comparison reveals differences primarily at the boundary regions of the sample. These discrepancies occur because the hybrid solver did not reach the global minimum of the QUBO model. The target minimum value derived from Fig. 2b is $-5,393,765.588239735$, whereas the minimum found by the hybrid solver is $-5,373,756.427596$. This suggests that the solver converged to a near-optimal solution instead of the global minimum, resulting in slight reconstruction variations at the sample boundaries.

### 2) Reconstruction Results for Sinograms with Limited Projection Angles

To demonstrate the superiority of our algorithm with smaller angular ranges, we set the maximum angle to be less than 180° in this experiment. The cross-sectional image of the tooth sample was sized at 50 × 50 pixels, with the axial level set to 15. To address concerns about using too few projection angles due to the smaller angular range, we initially divided the range from 0° to 180° into 100 angles and created 100 projection images. The sinogram generated from these 100 projection images is shown in Fig. 4a.

We first conducted an experiment using the first 50 projection images, corresponding to angles smaller than 90°. The image reconstructed using FFT with these 50 projection images is shown in Fig. 4b, where the pixel values deviate from the expected range of 0 to 1, and the boundary shapes are indiscernible. Similarly, Fig. 4c displays the result of applying FFT to 60 projection images, corresponding to angles smaller than 108°. Both images were reconstructed from their respective subsets of the sinogram shown in Fig. 4a. For reference, Fig. 4d presents the solution image obtained by applying FFT and thresholding methods to the entire sinogram.

When our quantum algorithm was applied to the same 50 projection images, the reconstructed image, shown in Fig. 4e, preserved the overall structure but exhibited significant errors in the top-left region. To enhance the image quality, we applied a post-processing method using pixel connectivity, yielding the refined image shown in Fig. 4f. To further assess the accuracy of our method, we overlaid this post-processed image with the solution image in Fig. 4d, producing the comparison image shown in Fig. 4g.

Next, we evaluated the impact of increasing the number of projection angles by applying our quantum algorithm to the 60 projection images, which produced the reconstruction shown in Fig. 4h. While the overall structure was preserved, errors remained visible in both the top-left and bottom-right regions. The post-processed result, after applying pixel connectivity, is presented in Fig. 4i, and the comparison with the solution image (Fig. 4d) is shown in the overlaid image in Fig. 4j. Comparing the overlaid images (Figs. 4g and 4j) clearly demonstrates that increasing the number of projection images from 50 to 60 reduces reconstruction errors and produces an image that more closely aligns with the solution image.

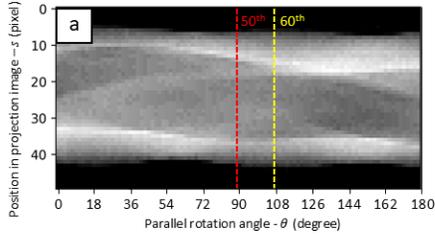

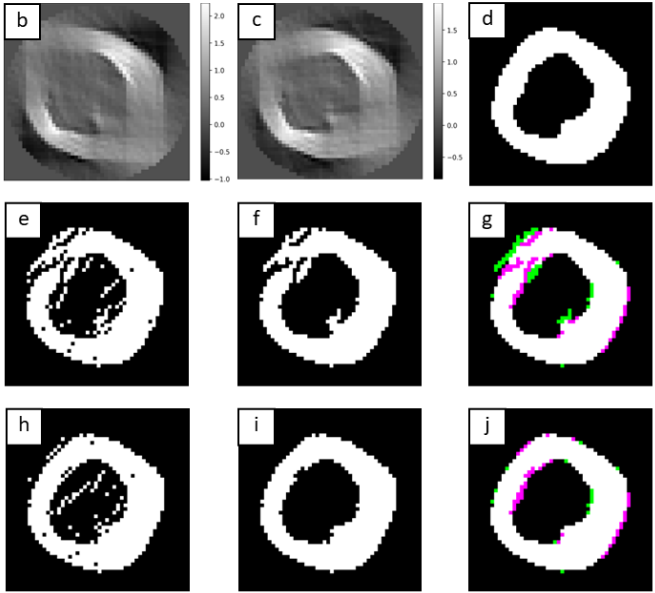

Fig. 4. Experimental results for limited angles on a 50 × 50 cross-sectional image at axial level 15 of the tooth sample. (a) Sinogram generated by dividing the range from 0° to 180° into 100 intervals. (b) Tomographic image reconstructed using FFT with the first 50 projection images from the sinogram in (a). (c) Tomographic image reconstructed using FFT with 60 projection images. (d) Solution image obtained by applying FFT to the entire sinogram and applying thresholding. (e-g) Results from our algorithm applied to the 50 projection images, including hybrid solver reconstruction, post-processing, and comparison with the solution image in (d). (h-j) Results from our algorithm applied to the 60 projection images, including hybrid solver reconstruction, post-processing, and comparison with the solution image.

To conduct an experiment on a noise-free sinogram, we also tested a 50 × 50 Shepp-Logan binary sample image with a zero-padding of thickness 11 added on all sides. As shown in Fig. 5a, we restricted the projection angles to 90° and used only 25 projection images, which is half of the total 50. The result of applying FFT to these 25 projection images, shown in Fig. 5b, reveals that the outer boundaries in the top-left and bottom-right corners, as well as the right ellipse among the two inner ellipses, are difficult to discern. In contrast, the result of applying our algorithm to this case, shown in Fig. 5c, perfectly matches the original sample image used in the experiment.

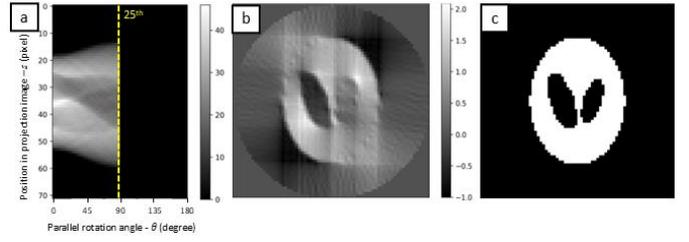

Fig. 5. Experimental results on limited angles for a 50 × 50 Shepp-Logan binary sample image with zero-padding. (a) Sinogram generated using the first 25 projection images (under 90°) out of 50, evenly divided from 0° to 180°. (b) Tomographic image reconstructed using FFT from the sinogram in (a). (c) Tomographic image reconstructed using our algorithm from the sinogram in (a), also representing the original sample image used in the experiment.

### B. Comparison of optimization performance between classical and quantum algorithms

The proposed method reconstructs tomographic images by formulating a QUBO model based on the regions of the sinogram unaffected by errors and then determining the minimum value of this QUBO model along with its corresponding solution. To demonstrate quantum advantage, we compare the results obtained from D-Wave's hybrid solver with those derived from traditional optimization tools applied to the same QUBO model.

All experimental results were evaluated based on the performance of each method, using the Mean Absolute Error (MAE) and the time required to compute the solution.

#### 1) Reconstruction Results for Sinograms with Ring Artifact-Inducing Errors

A Body tomographic image of size 50 × 50, shown in Fig. 1h, was zero-padded by 10 pixels on each side. Using this image, 50 projection images were generated at angles evenly spaced from 0° to 180°, forming the sinogram. To introduce errors, 10%, 20%, 30%, and 40% of the 50 positions (from position 11 to 60) in the sinogram were randomly selected, and artificial errors were introduced by multiplying the pixel values at these positions by random scaling factors. The sinograms with these artificially introduced errors are shown in the first column of Fig. 6. These horizontal strip errors, which are perpendicular to the position axis, result in circular artifacts when the tomographic image is reconstructed using FFT. As shown in the second column of Fig. 6, higher error rates lead to more prominent circular noise in the reconstructed tomographic images, making internal structures increasingly indistinguishable.

Next, we generate a QUBO model to apply our algorithm. The pixel values of the tomographic image to be reconstructed are represented as described in (16). Since the pixel values in the images used in the experiments are limited to 0, 1, 2, and 3, each pixel can be expressed as a combination of qubits using (16), such that $I_{ij} = q_1 + q_2 + q_3$. Additionally, pixel values at positions affected by errors in the sinogram are excluded during the QUBO model generation. When the generated QUBO model is solved using D-Wave's hybrid solver, as shown in the last column of Fig. 6, the solution is successfully obtained for all error cases. However, solving the same QUBO model using Gurobi or SA reveals that most solutions cannot be obtained, except for the case of a sinogram with 10% error solved by Gurobi. Furthermore, as seen in the third and fourth columns of Fig. 6, higher error rates result in reconstructed images that deviate significantly from the original image. This finding is further supported by Table I, which shows that, apart from one case, all experiments using other classical methods failed to find a solution. Additionally, the computational time required to find a solution highlights the efficiency of the hybrid solver. On average, the hybrid solver takes 27 s, whereas SA requires an average of 86 s, approximately three times longer. Gurobi's presolve results, on average, take 218 s, making it approximately eight times slower than the hybrid solver.

To investigate cases where more qubits are used to represent a single pixel, a head tomographic image of size $30 \times 30$, shown in Fig. 1k, was zero-padded by 5 pixels on each side. Using this image, 30 projection images were generated at angles evenly spaced from 0° to 180°, forming the sinogram. To introduce errors, 10% of the 30 positions in the sinogram (positions 6 to 35) were randomly selected, and artificial errors were generated by multiplying the pixel values at these positions by random scaling factors. In this setup, each pixel of the tomographic image was represented using 10 qubits, such that $I_{ij} = q_1 + q_2 + \cdots + q_{10}$. The results of experiments conducted under the same conditions as the previous examples are shown in Fig. 7 and Table I. As observed, neither Gurobi nor SA was able to find a solution in this case, while the hybrid solver successfully obtained the solution.

### 2) Reconstruction Results for Sinograms with Limited Projection Angles

In this section, we introduce a different type of error to the $50 \times 50$ Body tomographic image and its corresponding sinogram used in the previous section by randomly deleting projection images at certain angles. The proportions of randomly selected angles tested were 10%, 30%, and 50%. Sinograms with these errors are shown in the first column of Fig. 8.

For the 10% error case, Gurobi, SA, and the hybrid solver all successfully found the correct solution. To evaluate the FFT results using the iradon function, we excluded portions of the sinogram and projection angles containing errors. The FFT-reconstructed image was then rounded to integer pixel values for consistency with the original image format. As shown in Fig. 8, the FFT reconstruction performs well for the 10% error case, with differences from the other three methods appearing in only a few pixels. Additionally, as indicated in Table I, FFT achieves the lowest MAE of 0.0180 across all experiments. However, as the number of missing projection images increases, the FFT-reconstructed images show noticeable noise and distortions, which degrade the overall quality of the reconstruction.

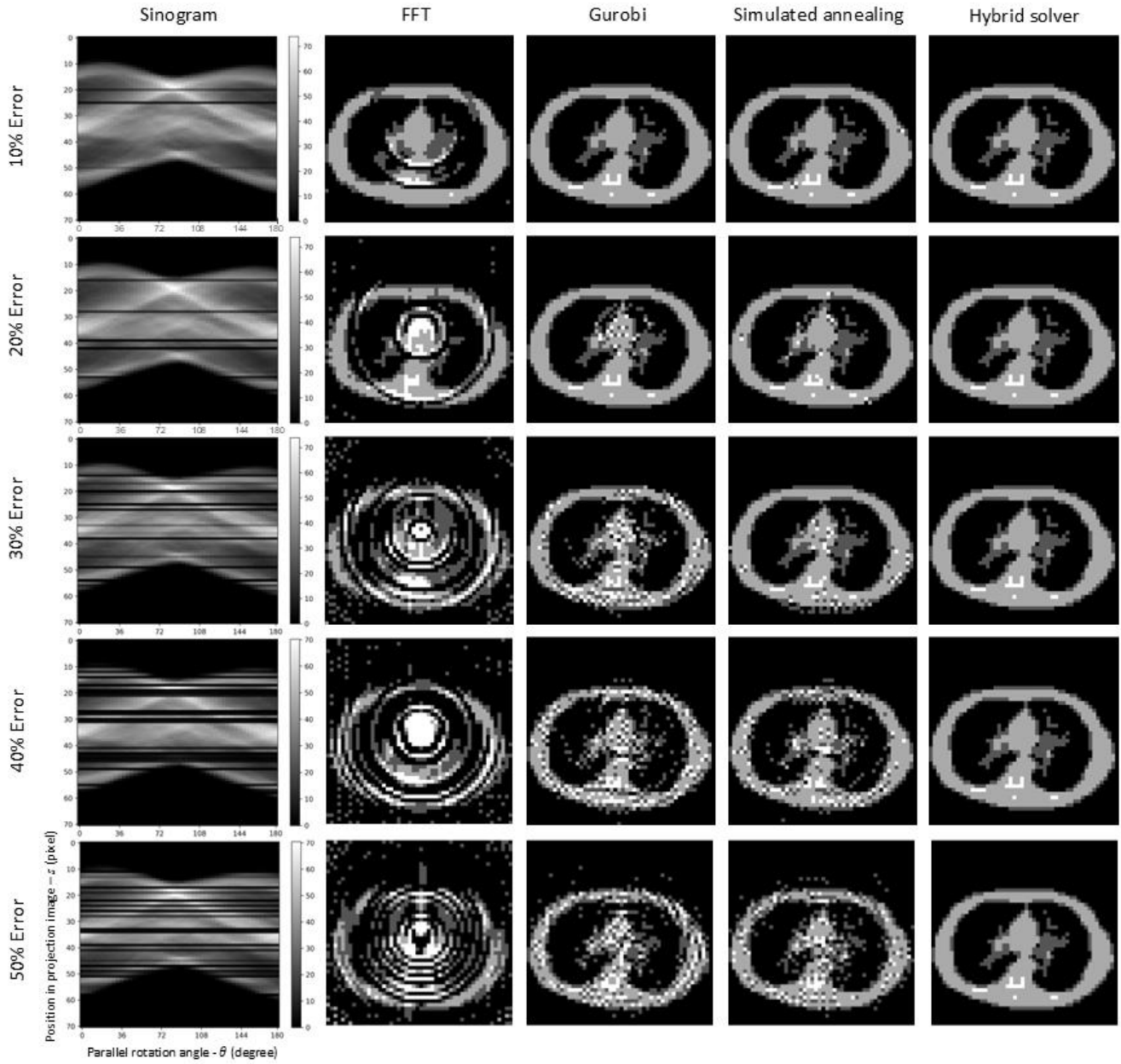

Fig. 6. Sinograms with different error rates, leading to ring artifacts, and the corresponding tomographic images reconstructed using four methods: FFT, Gurobi, SA, and the Hybrid Solver.

TABLE I

SUMMARY OF EXPERIMENTAL PARAMETERS AND RESULTS, INCLUDING THE IMAGE SIZE, PIXEL VALUE RANGE, ERROR TYPE, ERROR RATE, MEAN ABSOLUTE ERROR (MAE) FOR EACH METHOD, AND THE COMPUTATION TIME REQUIRED BY EACH METHOD.

| Image Size | Range of Pixel Value | Error Type | Error Rate | MAE | | | | Time (seconds) | | |
|---|---|---|---|---|---|---|---|---|---|---|
| | | | | FFT | Gurobi | SA | Hybrid | Gurobi | SA | Hybrid |
| $50 \times 50$ | [0, 3] | Ring artifact | 10% | 0.1172 | 0 | 0.0032 | 0 | 203.16 | 103.72 | 27.30 |
| $50 \times 50$ | [0, 3] | Ring artifact | 20% | 0.2724 | 0.0268 | 0.0200 | 0 | 217.85 | 86.73 | 27.25 |
| $50 \times 50$ | [0, 3] | Ring artifact | 30% | 0.5688 | 0.1160 | 0.0412 | 0 | 237.29 | 65.46 | 27.20 |
| $50 \times 50$ | [0, 3] | Ring artifact | 40% | 0.7244 | 0.1676 | 0.1084 | 0 | 258.56 | 57.71 | 27.30 |
| $50 \times 50$ | [0, 3] | Ring artifact | 50% | 0.6976 | 0.1816 | 0.1496 | 0 | 246.82 | 48.04 | 27.29 |
| $30 \times 30$ | [0, 10] | Ring artifact | 10% | 1.2778 | 0.3556 | 0.3056 | 0 | 177.60 | 179.03 | 34.82 |
| $50 \times 50$ | [0, 3] | Limited angle | 10% | 0.0180 | 0 | 0 | 0 | 244.86 | 88.24 | 27.25 |
| $50 \times 50$ | [0, 3] | Limited angle | 30% | 0.0684 | 0.0960 | 0.0200 | 0 | 211.85 | 86.16 | 27.23 |
| $50 \times 50$ | [0, 3] | Limited angle | 50% | 0.2368 | 0.1672 | 0.1212 | 0 | 240.11 | 64.38 | 27.25 |

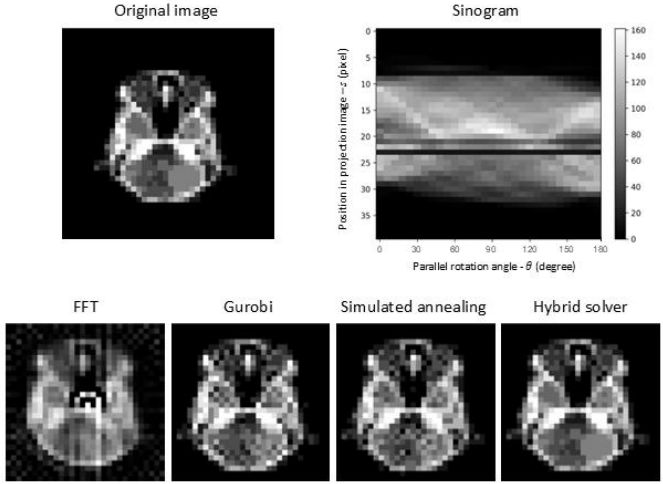

Fig. 7. Head tomographic image reconstruction results and the corresponding sinograms using four methods: FFT, Gurobi, SA, and Hybrid Solver. The sinogram without errors and its corresponding reconstructed tomographic image are shown in Fig. 1l and Fig. 1k, respectively.

As shown in Table I, the hybrid solver outperformed the other methods in terms of computation time, as observed in the previous example. It successfully found the solution for all three error cases (10%, 30%, and 50%) in approximately 27 s. In contrast, FFT, Gurobi, and SA were unable to recover the solution for the 30% and 50% error cases. This highlights the superior robustness and efficiency of the hybrid solver in handling sinograms with higher levels of missing projection data.

## V. Discussion

Quantum optimization-based tomographic image reconstruction algorithms offer faster and clearer results compared to classical methods [17]. In this study, we demonstrate further quantum superiority in reducing the angular range of projection images used for tomographic reconstruction and in artifact removal within the reconstructed tomographic images.

First, we address the projection angles used in the quantum algorithm. Traditionally, 180° projections are employed for tomographic image reconstruction, while 360° projections are sometimes used with misaligned projections to produce clearer images [27]. In contrast, quantum algorithms can achieve the highest-quality tomographic image reconstructions using projections of just up to 90 degrees. The novel quantum algorithm exhibits greater consistency than classical methods, which typically utilize first-order optimization functions. By employing a second-order optimization function for each pixel of the sinogram and its projected pixels in the superposition state, the quantum algorithm enables clearer tomographic image reconstruction across all types of tomography. This capability, in particular, addresses the issue of missing wedges, a common challenge in electron tomography [28]. This advancement will not only significantly enhance electron tomography but also improve the efficiency of data acquisition across all tomographic systems.

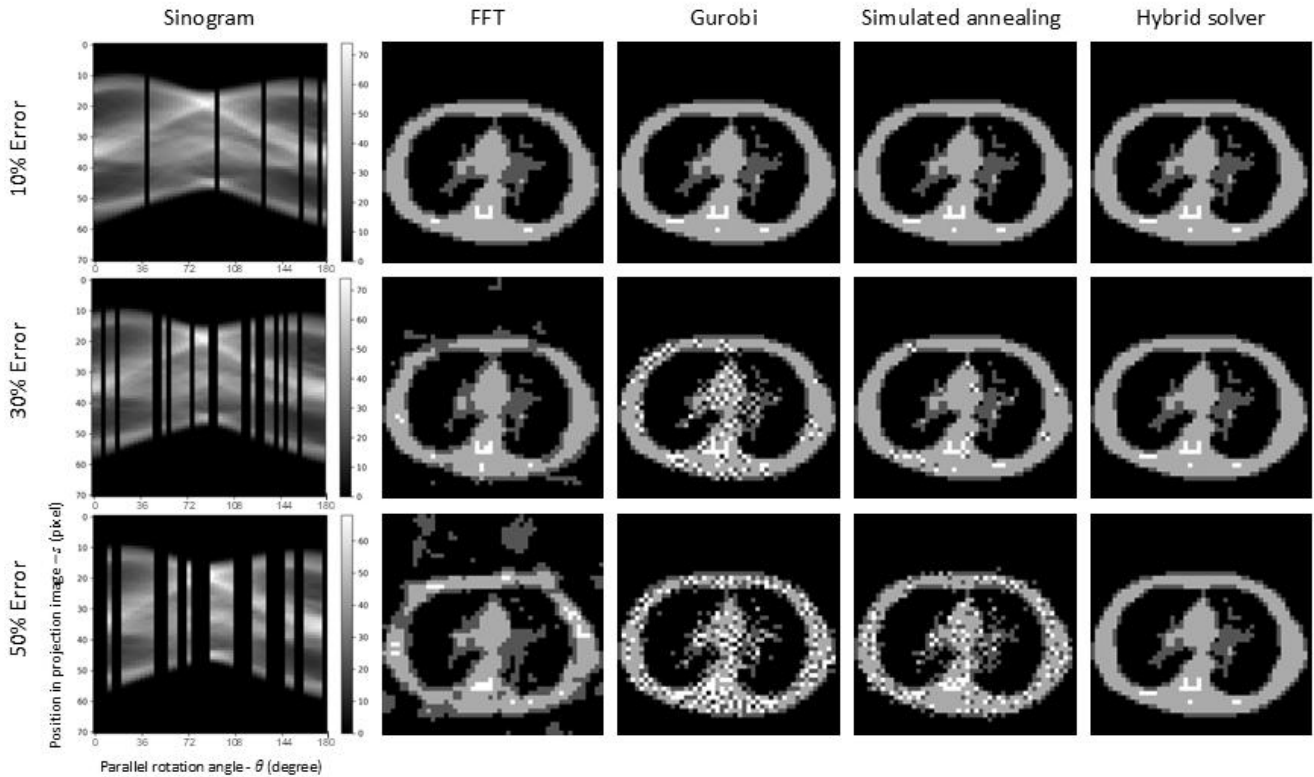

Fig. 8. Sinograms with missing projection images at randomly selected angles, corresponding to varying error rates, generated from the Body tomographic image, and the corresponding tomographic images reconstructed using four methods: FFT, Gurobi, SA, and Hybrid Solver.

When a sinogram containing horizontal error regions is reconstructed using classical algorithms, ring artifacts typically appear in the resulting image. In contrast, our algorithm prevents this issue by focusing on each pixel of the image to be reconstructed, converting them into qubits, and generating the QUBO model. This approach allows for the exclusion of unreliable regions in the sinogram while still leveraging the qubit information from the remaining reliable areas. Consequently, even when unreliable regions are excluded, the qubit information from the reliable regions continues to contribute to the reconstruction process.

To validate our method, we conducted experiments by injecting artificial errors into the sinogram to simulate the creation of ring artifacts. This approach was necessitated by the current limitations in the number of qubits available on quantum computers, which require resizing the image to a smaller dimension for processing. During resizing, errors can become indistinguishable from surrounding pixel values, leading to the loss of error-specific characteristics. To ensure a controlled and reproducible evaluation, we resized the images to the experimental dimensions and introduced artificial errors into the sinograms, generating a consistent testing environment.

Additionally, the pixel values of the tomographic image to be reconstructed were set as integers. Among the various experimental datasets, the tooth sample was reconstructed using real X-ray data. Since the tooth sample consists of a single MAC, preprocessing was conducted to assign a value of 1. Even after removing 50% of the data from the sinogram and applying our algorithm, the $100 \times 100$ cross-sectional tooth image was successfully reconstructed. The reconstructed result closely matched the image obtained by applying FFT to the complete, error-free sinogram, except for certain boundary regions of the sample. However, it is important to note that FFT-based reconstruction also relies on a thresholding technique, and adjustments to the threshold value can influence the reconstruction at the boundaries. Considering this, our algorithm demonstrates a clear advantage over classical reconstruction methods by producing high-quality results even when only half of the data is available.

Using our algorithm, the $50 \times 50$ Body tomographic image, where each pixel consists of three integers, achieved perfect reconstruction results, with an MAE of 0, even under up to 50% error in the sinogram. Similarly, the $30 \times 30$ Head tomographic image, where each pixel consists of ten integers, demonstrated perfect reconstruction with an MAE of 0 despite up to 10% error in the sinogram. However, when the error level was increased to 20%, the hybrid solver failed to reach the global minimum. For the given QUBO model, the target minimum value is -4,607,247.045108841, while the minimum found by the hybrid solver was -4,607,144.206576, indicating that the solver could not locate the exact optimal solution. This result suggests that while our algorithm remains highly effective at lower error levels, the current performance limitations of hybrid solvers prevent the achievement of perfect reconstruction under higher noise conditions. While our algorithm shows significant promise, the current state of quantum computing limits its scalability and applicability in real-world tomographic image reconstruction scenarios. For instance, the number of qubits available on current quantum processors restricts the size of the sinogram data that can be processed, which limits the algorithm's ability to handle larger and more complex tomographic datasets typically used in clinical or industrial applications. Nevertheless, as quantum annealing technology advances, the ability to handle larger QUBO models with more qubits is expected to improve, potentially enabling accurate reconstruction even in cases with greater sinogram error.

Given the limited material groups (air, soft tissue, bone) in typical tomographic images, and assuming knowledge of their respective MACs, we estimate that 3–4 qubits per pixel, in conjunction with (11) and (16), would be sufficient for accurate reconstruction. In this scenario, for a $50 \times 50$ image size with data containing up to 50% error, we can still expect accurate tomographic image reconstruction using our algorithm.

In all experiments involving the Body and Head tomographic images, the results obtained using our algorithm with the hybrid solver deviated from the target minimum value of the QUBO model by values on the order of $10^{-6}$ to $10^{-8}$, which are negligible and do not affect the validity or interpretation of the results. This indicates that the algorithm successfully identified solutions very close to the target minimum. In contrast, as the error level increased, other tools such as Gurobi and SA, which also utilized the same QUBO model, failed to find a valid solution. Furthermore, the classical reconstruction method, FFT, also failed to accurately reconstruct the tomographic images in these experiments. These comparative results strongly demonstrate the quantum supremacy of our algorithm for tomographic image reconstruction on quantum computers. Furthermore, when projections containing metal density are used, the internal structures, excluding the metal within the sample, can be clearly reconstructed by applying the quantum algorithm to projections that exclude pixels with metal density.

These findings underscore the significant advantages of quantum algorithms over classical methods in tomographic image reconstruction, particularly under conditions with incomplete or erroneous data. This highlights the potential of quantum computing as a transformative tool for tomographic imaging systems. Our future research direction will focus on the structural analysis of nano catalysts using new algorithms and data from high-resolution transmission electron microscopy. We will also continue to study the possibility of utilizing the fan-beam CT system for structural analysis of samples and the production of related medical devices using precision measurement and precise quantum 3D images in the medical imaging field.

## DATA AVAILABILITY

The Python code used in this paper can be found on the author's GitHub site (https://github.com/ktfriends/Quantum_CT_reconstruction/tre

e/main/Quantum_Supremacy).

ACKNOWLEDGMENT

Synchrotron data was conducted at beamline 6C of the Pohang Light Source-II, which is funded by the Ministry of Science and ICT (MSIT) and POSTECH in Korea. This research was supported by the quantum computing technology development program of the National Research Foundation of Korea (NRF) funded by the Korean government (Ministry of Science and ICT (MSIT)) (No. 2020M3H3A111036513). H. L. is supported by the National Research Foundation of Korea(NRF) grant funded by the Korea government(MSIT)(RS-2024-00352408). K. J. is supported by the MSIT(Ministry of Science and ICT), Korea, under the ITRC(Information Technology Research Center) support program(IITP-RS-2024-00437284) supervised by the IITP(Institute for Information & Communications Technology Planning & Evaluation).)